\documentclass [12pt]{article}
\usepackage{amsmath,epsfig}
\def\4qc{\rm \langle\langle 0|{\bar q}\Gamma t^a q{\bar q}\Gamma t^a q|0
\rangle\rangle}
\def\qq{\langle\langle\rm{\bar q q}\rangle\rangle}
\def\sigm{t_\sigma}

\def\g5{\gamma^5}

\def\d4k{{d^4k\over (2\pi)^4}}

\newcommand{\beq}{\begin{eqnarray}}
\newcommand{\eeq}{\end{eqnarray}}
\newcommand{\beqno}{\begin{eqnarray*}}
\newcommand{\eeqno}{\end{eqnarray*}}
\def\lsim{\mathrel{\rlap{\lower4pt\hbox{\hskip1pt$\sim$}}
    \raise1pt\hbox{$<$}}}         
\def\gsim{\mathrel{\rlap{\lower4pt\hbox{\hskip1pt$\sim$}}
    \raise1pt\hbox{$>$}}}         

\begin{document}
%
\title{Light-Quark Mesons and Four-Quark Condensates at Finite Temperature}
\author{Mikkel B. Johnson\\
       Los Alamos National Laboratory, \\
       Los Alamos, New Mexico\\
        Leonard S. Kisslinger\\
        Department of Physics,\\
       \vspace{5mm}
       Carnegie Mellon University, Pittsburgh, PA 15213}
\maketitle
\indent
\abstract{We propose an analog of the familiar gap equation for the case of 
four-quark 
condensates at finite temperature.  The condensates of interest correspond to 
scalar, vector, psudoscalar, axial vector, and tensor Dirac structures.  
Working with correlators at zero chemical potential without factorization,
we arrive at coupled equations for these four-quark condensates
and the masses of certain light-quark mesons.  We study the temperature
dependence of the four-quark condensates and masses; in one 
of our models, factorization of the four-quark condensates is shown to be 
increasingly violated as the temperature is increased toward $T_c$.  The
2$^{++}$ tensor mesons a$_2$(1320)-f$_2$(1270) are identified as especially
sensitive probes of the four-quark condensates.}
\newpage
\section{Introduction}
\hspace{.5cm}

The early universe is believed to have evolved through a stage
beginning about 100 ps and ending about one $\mu$ sec after the big bang, 
during which time leptons had acquired their masses and strongly interacting 
matter existed primarily as a plasma of quarks and gluons. 
As the universe further expanded and cooled, when the temperature
reached the value $T = T_c \simeq 200$ MeV matter began to undergo a phase
transition as quark condensates and possibly hadrons began
to form. Upon further expansion and cooling, this transition ran to 
completion with all quarks and gluons then having combined into hadrons. 

Our paper concerns the physics that developed as the universe passed from 
$T \simeq T_c$ to $T=0$.  The relevant issues are identified by focusing on 
mesons in such a medium.  A large body of work on vector mesons at finite T 
already exists, having been motivated by the fact that vector mesons can 
couple directly to photons providing an experimental lepton-pair signal of the 
masses and widths of these particles.  However, without the factorization 
assumptions upon which
most of this work relies, and which we describe in detail below, all the
mesons become coupled through four-quark condensates. In the present paper we 
describe this physics without factorization, deriving a set of coupled 
equations for the meson masses and for the relavant four-quark condensates, 
and giving solutions with various scenarios for the finite-T spectral densities 
and two- and four-quark condensates.
     
The quantitative characterization of this sequence of events lies beyond the 
capabilities of theoretical analysis at present.  However, because of its
high interest, experimental programs have been planned to create similar
conditions in the laboratory.  It is the explicit aim for the RHIC 
(Relativistic Heavy Ion Collider) project at Brookhaven National Laboratory
(BNL) to initiate a transition to matter at $T\approx 200$ 
MeV or greater, the quark-gluon plasma, within a large enough volume,
and for a sufficiently extended time, to allow the study of its properties.  
This is envisioned to occur when the matter of large nuclei is compressed
and heated as a result of violent, head-on collisions between nuclei that
have been accelerated to relativistic energies and allowed to collide in the 
RHIC storage ring.  Encouraging evidence from preliminary experiments at 
the AGS at BNL and the SPS at CERN that a hot, dense fireball is formed in 
relativistic heavy ion collisions~\cite{ms} keeps interest high in 
the RHIC program.  The experimental determination of the nature of 
this transition is expected to guide the development of theoretical methods to 
treat the same physics.  
     
In this paper we will develop analytical tools capable of addressing
the temperature dependence of the four-quark condensates characterizing
the quark-gluon plasma that may be created in this fashion.
We are particularly interested in the four-quark condensates because they 
are fundamental quantities that characterize the nonperturbative vacuum 
in quantum chromodynamics (QCD) and are needed within the framework of QCD
sum rules~\cite{svz} to 
describe properties of both mesons (for a review, see Ref.~\cite{rry1})
and baryons~\cite{i}.  They are especially important in the case of 
the light-quark mesons, which are the lowest excitations of the physical vacuum.
The seminal work on this problem at $T\neq 0$ was done by 
Bochkarev and Shaposhnikov, Ref.~\cite{bs}.  They investigated an important 
four-quark condensate, namely that contributing to the mass of the 
$\rho$ meson, using QCD sum rules extended to finite temperature. 

Bochkarev and Shaposhnikov introduced a dynamical model for the temperature 
dependence of this four-quark condensate, thus avoiding the so-called 
factorization approximation, according to which each four-quark condensate
becomes trivially proportional to the square of the familiar quark 
condensate, assuming saturation by the vacuum.  A number of other 
authors have also calculated the mass of the 
rho meson using QCD sum rules at finite $T$~\cite{dn,fhl}, generally using 
factorization. For the related physics of hadrons at $T=0$ but finite density
it has also been seen~\cite{fb} that the factorization of the four-quark
condensates is a major question, and that, e.g., in nuclear
matter the theoretical result for the nucleon mass is criticaly dependent 
on the factorizaton assumption. In a study of the $\Delta(1236)$ in
nuclear matter it has been shown that factorization 
of four-quark condensates is not valid for hadrons in a medium of non-zero 
density~\cite{jk}.  At zero baryon density there is evidence that it is 
a bad approximation~\cite{gbp}; however, because it is simple and convenient 
the approximation
is widely used.  To our knowledge, the present work is the first attention 
given to the study of four-quark condensate in dynamical models 
since the original work of Bochkarev and Shaoposhnikov.  

An important set of four-quark condensates are those of the form 
\beq
{\hat Q}^\Gamma \equiv \langle 0|:{\bar q}\Gamma t^a q{\bar q}\Gamma t^a q:|0\rangle
\label{qqqq}
\eeq
where $t^a$ are the SU(3) Gell-Mann color matrices and $\Gamma$ corresponds
to the set of five independent Dirac structures $\Gamma=1$, 
$\gamma_5$, $\gamma_\mu$,$\gamma_5\gamma_{\mu}$, and $\sigma_{\mu\nu}$, 
with an implied contraction over the Dirac indices.  It is particularly
relevant for the present work that these condensates make major contributions 
to the masses of the light-quark mesons generated by 
the corresponding five quark currents, 
\beq
J^\Gamma(x) & = &{\bar q(x)} \Gamma q(x) ,
\label{current}
\eeq
where q(x) are the u,d-quark fields contracted over their color labels.
The mesons generated by a particular current $J^\Gamma(x)$ 
are those that are excited from the vacuum by the action of the current, 
i.e., those that satisfy 
\beq
 <0|J^\Gamma (0)|meson(\Gamma )> & = & g_\Gamma \Psi_\Gamma ({\vec p})\neq 0,
\label{eq-eta}
\eeq
where $\Psi_\Gamma ({\vec p},\lambda)$ is a wave function characteristic of the
meson and $g_\Gamma$ is a structure parameter.  We will often refer to the 
mesons and other quantities associated with the five different Dirac 
$\Gamma$ by {\it s, ps, v, av}, and $\sigm$, respectively.   

In our discussion of the mesons, we will also be interested in the axial, 
tensor and axial tensor mesons because their masses depend on the same set
of five four-quark condensates.  The currents required for these 
mesons~\cite{rry1} are somewhat more complicated than the forms given in 
Eq.~(\ref{current}).  One should also note that to determine the mass
in the scalar (0$^{++}$) case, the current must include scalar glueball mixing;
this is discussed in detail below. The most important mesons associated with 
each current will be specified below.  

\section{QCD Sum Rules for Mesons}
\hspace{.5cm}

The use of QCD sum rules for our purpose is suggested because this
framework makes a relatively simple and explicit connection between the 
four-quark condensates and light meson masses.  QCD sum rules have a number
of drawbacks, however, such as uncertainty in how to treat the continuum,
which becomes more and more serious issue as the temperature is raised.  For 
this and other reasons, we do not aim to make predictions of QCD but rather
develop QCD sum-rule based models, in which parameters are introduced to
account for the temperature dependence of the continuum.  We find our 
results reasonable, and we believe our {\it ad hoc} assumptions are capable of 
being improved as experimental results warrent it.
   
\subsection{General Considerations}
\hspace{.5cm}

In the method of QCD Sum Rules~\cite{svz}, the propagator of a hadron is 
represented by a two-point function, usually called a correlator, defined as
\beq
\Pi^\Gamma(p) & = 
  & i\int d^4x e^{ix\cdot p}<0|T[J^\Gamma(x)J^\Gamma(0)]|0>,
\label{eq-pi}
\eeq
where for light-quark mesons the currents $J^\Gamma(x)$ are given 
in Eq.~(\ref{current}) and other forms specified below.  This 
differs from the two-point function of standard field theories in that the 
$J^\Gamma(x)$'s are composite.  Similar correlators are introduced in the 
study of baryons, but in the present work we focus only on the mesons for 
reasons discussed above.

A QCD sum rule for a quantity of interest is possible when the correlator
is saturated by hadronic states, i.e. by using Eq.~(\ref{eq-eta}) in 
Eq.~(\ref{eq-pi}), producing a dispersion relation in which
the quantity of interest appears explicitly.  The expression resulting from
this method of evaluation is often called the right-hand side (R.H.S.) of the
sum rule.  One form for the dispersion relation, used in the present work,
is:
\beq
\Pi^\Gamma_{R.H.S.}(p)=\int^\infty_{-\infty}\frac{\rho(u,{\vec p})du}
{u-p_o+i\eta}
\label{rhs}
\eeq
with
\beq
\rho(E,{\vec p})=\sum_{\lambda\alpha}\int\frac{d^3{\vec p}_\alpha}{(2\pi )^3} 
\langle 0|
J^\Gamma|m_\alpha ({\vec p},\lambda)\rangle\langle m_\alpha ({\vec p},\lambda)|
J^\Gamma|0\rangle\frac{(2\pi)^3\delta({\vec p}-{\vec p}_\alpha)}{2E_\alpha}
\delta(E-E_\alpha) ,
\label{imrho}
\eeq
where $|m_\alpha({\vec p},\alpha)\rangle$ of 4-momentum $(E_\alpha,{\vec p})$
is a complete set of intermediate states of quantum numbers $\lambda$,
including the continuum.  

To make a connection to QCD, the correlator is then evaluated microscopically 
by expanding in operators of increasing dimension using an operator product 
expansion (o.p.e).  The resulting evaluation is often called the left-hand
side (L.H.S.).  For the quark propagator in space-time,
\beq
 S_q(x) & = & <0|T[q(x)\bar{q}(0)]|0>, 
\nonumber\\
 	& = & S_q^{P.T.}(x) + S_q^{N.P.}(x),
\label{eq-prop}
\eeq
with
\beq
 S_q^{N.P.}(x) & = & <0|:q(x)\bar{q}(0):|0>,
\label{eq-npprop}
\eeq
where $S_q^{P.T.}(x)$ is the usual perturbative quark propagator, illustrated 
in Fig.~1a, and $S_q^{N.P.}(x)$ is the nonperturbative quark propagator, which 
\begin{figure}
\begin{center}
\epsfig{file=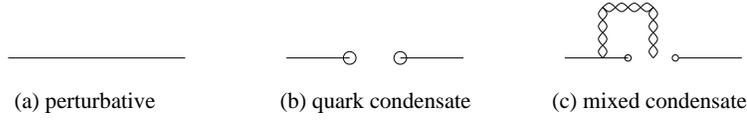,width=10cm}
\caption{Quark propagator.}
{\label{fig.1}}
\end{center}  
\end{figure}             
vanishes in the perturbative vacuum.  The operator product expansion gives
\beq
  S_q^{N.P.}(x)  & = & \frac{-1}{12}<0|:\bar q q:|0> + \frac{x^2}{3 \cdot 2^6}
 <0|:\bar q \sigma\cdot G q:|0> + ...
\label{eq-snp}
\eeq
in terms of the quark condensate, whose value is given by
\beq
\langle 0|: {\bar q}q:|0\rangle=-(0.25~{\rm GeV})^3 ,
\label{2qc}
\eeq
a mixed condensate, and so forth as
illustrated in Figs. 1b, 1c.  The justification for the o.p.e. is that the 
evaluation of $\Pi^\Gamma (p)$ is done using a Borel transformation with the 
Borel mass in the range of about 1 GeV, allowing a short-distance expansion.  
 
The microscopic derivation of the correlator $\Pi^\Gamma (p)$ is completed by 
using 
the expansion of S$_q$(x) given in Eqs.~(\ref{eq-prop},\ref{eq-snp}), leading
to the perturbative term illustrated in Fig.~2a.  To this is added
\begin{figure}                                                                
\begin{center}
\epsfig{file=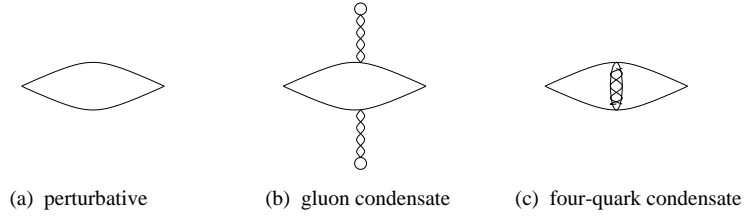,width=10cm}
\caption{O.P.E. for meson correlator.}                                                          
{\label{Fig.2}}           
\end{center}
\end{figure}
terms with gluons attached to two or more quark propagators, with the
dimension four gluon condensate term illustrated
in Fig.~2b.  Finally, there are important contributions that come from the 
four-quark terms illustrated in Fig.~2c and that are the focus of this work.  
For light-quark mesons, the quark condensate terms are
suppressed because they are multiplied by a small number, the mass of an u or d 
quark.  Consequently, the most important nonperturbative contributions 
to the light-meson masses are four-quark condensates Fig.~2c, and the gluon 
condensate, Fig.~2b.

We will denote the four-quark condensates corresponding to the
correlator $\Pi^\Gamma$ as ${\hat{\hat Q}}^\Gamma$.  These have the following
form:  ${\hat{\hat Q}}^v\equiv {\hat Q}^{av}+\frac{2}{9}|J_0|^2$, 
${\hat{\hat Q}}^{av}\equiv {\hat Q}^v+\frac{2}{9}|J_0|^2$, 
${\hat{\hat Q}}^s\equiv {\hat Q}^{\sigm} +\frac{2}{3}|J_0|^2$, 
${\hat{\hat Q}}^{\sigm}\equiv {\hat Q}^s+\frac{1}{36}|J_0|^2$ and
${\hat{\hat Q}}^{ps}\equiv {\hat Q}^{a\sigm}+\frac{2}{3}|J_0|^2$, where 
${\hat Q}^\Gamma$ is defined in Eq.~(\ref{qqqq}) and
\beq
|J_0|^2\equiv \langle 0|:{\bar q}\gamma_\mu t^aq \sum_f
{\bar q_f}\gamma^\mu t^aq_f:|0\rangle .
\label{qhat}
\eeq
As the four-quark condensates are mediated by gluon exchange, they 
contribute to the correlator proportional to $\alpha_s=g_s^2/4\pi\approx 0.7$
at a scale of $\mu=200$ MeV~\cite{svz}, where $g_s$ is the strength of
the quark-gluon coupling.  Note that we have given four-quark condensates 
corresponding to correlators for all five Dirac structures, including 
$\Pi^{\sigm}$, and that {\it a}$\sigm$ refers to the 
current $J^\Gamma$ with $\Gamma=\gamma_5\sigma_{\mu\nu}$.  Other four-quark 
condensates are needed for the correlators of the axial, tensor and
axial tensor mesons, and these are defined below.

As we have noted, the square of the quark condensates instead of 
${\hat {\hat Q}}^\Gamma$ appears in versions of the theory
when the {\it factorization} approximation is applied to the four-quark 
condensates.  This approximation has been successful in describing 
the properties of hadrons at $T=0$ in free space (see, however, 
Ref.~\cite{gbp}).  
Since factorization may not be valid when $T\neq 0$, in developing our 
ideas we will try to avoid it as far as possible.  Consequently, we will 
generally be using the exact representation of the four-quark condensates as 
given in Fig.~2c. At $T=0$, the gluon condensate has the familiar value
\beq
g_s^2\langle 0|:G^a_{\mu\nu}G^{a\mu\nu}:|0\rangle \equiv \langle g_s^2 G^2\rangle
\simeq 0.47~{\rm GeV^4} .
\label{glue}
\eeq

\subsection{Extension to Finite Temperature} 
\hspace{.5cm}

We want to investigate the excitation modes of hadronic matter brought into 
contact with a heat bath.   Since we assume temperature equilibrium, 
the heat bath is the local environment for 
describing the properties of mesons propagating in the 
early universe or in the quark-gluon plasma created in the aftermath of 
colliding relativistic heavy ions in RHIC experiments.  The heat 
bath elevates the temperature of all matter in the vacinity of the meson in 
question to temperature $T$, including the constituents of the vacuum.

Statistical mechanics allows us to calculate the thermal average 
$\Pi^\Gamma(p,T)$,
\beq
\Pi^\Gamma(p,T) & = 
  & i\int d^4x e^{ix\cdot p}\theta(x_0)\langle\langle 0|J^\Gamma(x)
J^\Gamma(0)]|0\rangle\rangle ,
\label{eq-pit}
\eeq
where $p=(\omega,{\bf p})$ and where $\langle\langle ... \rangle\rangle$
stands for the Gibbs average.  As in the case of the corresponding operator 
given in Eq.~(\ref{eq-pi}), the poles of this operator correspond to the 
excitation spectrum of the medium.  Note that at finite temperature, one works
with the retarded rather than the causal propagator.

The quantity in Eq.~(\ref{eq-pit}) may be evaluated in field theory using the 
Matsubara formalism~\cite{m}, which recognizes the similarity between the time 
evolution operator introduced in field theory at $T=0$ and the Grand canonical 
ensemble introduced at $T\neq 0$.  The relevant formal distinction is the 
replacement of ${\it time}\rightarrow{\it i/T}$ to obtain the $T\neq 0$ 
results from the $T=0$ results.  Since the limits of integration are finite and bounded by $T$
at $T\neq0$, a boundary condition must be specified that is not needed in the
case of the case of field theory at zero temperature.  These differences
translate into the following rule for transforming a Feynman diagram 
at $T=0$ to a corresponding diagram at finite temperature:  the momentum 
$p^\mu\rightarrow (i\omega_n, {\bf p})$, where
\beq
\omega_n&\rightarrow &2\pi Tn~~~~~~~~~~~~{\rm bosons~(gluons),~and}\nonumber\\
\omega_n&\rightarrow &2\pi T(n+\frac{1}{2})~~~~{\rm fermions~(quarks)} ,
\label{mat}
\eeq
where $n$ runs over positive and negative integers (including zero).
Thus, instead of integrals over $\omega$ in Feynman diagrams, one has sums
over $n$.  When Feynman diagrams of a causal Green's function are translated 
in such a fashion, it can be shown that one 
recovers the Fourier transform of the advanced (retarded) real-time Green's 
function (at finite temperature $T$) when the analytical continuation 
$\omega_n \rightarrow i\omega+(-)\eta$ is made~\cite{ks}.

The Matsubara formalism may be applied to the correlator in Eq.~(\ref{eq-pi}) 
and hence to calculate the meson masses as a function of temperature using QCD 
sum rules.  One merely has to observe that the o.p.e. is made using Euclidian
variables, so the appropriate replacement of the Matsubara frequency is
now $\omega_n\rightarrow\omega$.  We will be interested in the excitation 
spectrum
at {\bf p}=0, so we will then be making the Borel transform (see Sect.~2.3)
on the variable $\omega$.  

In obtaining the QCD sum rule, one must recognize that the on the right-hand 
side, the residue at the pole of the advanced (or retarded) propagator is the 
familiar one except for the presence of the occupation probability factor 
(for bosons) $\coth(m_\Gamma/2T)$.  On the left-hand side, the perturbative 
quark term Fig.~2a is evaluated in accord with the above considerations in 
terms of the Matsubara propagator 
\beq
\frac{1}{i \omega_n\gamma_0-{\bf \gamma}\cdot {\bf p} - m} ,
\label{prop}
\eeq
and the nonperturbative terms are evaluated with the temperature-dependent
condensates.  

The temperature dependence of the quark condensate $\qq$ has been evaluated in
lattice gauge theory~\cite{boyd} and in the model of Nambu and 
Jona-Lasinio~\cite{hk} extended to finite temperature
using the Matsubara formalism.  In the latter picture, the two-quark condensate
drops toward zero at a temperature of about 160 MeV at zero chemical potential.
The models of the temperature dependence of the $\rho$ meson mass developed
in Refs.~\cite{dn,fhl} were developed using factorized four-quark condensates 
with $\qq$ having a temperature dependence of the general character found 
there.  The temperature dependence or the quark condensate has also be studied 
in instanton models~\cite{sur}.

As we have remarked, if one avoids the factorization approximation  then the 
actual contribution of $\qq$ to the light-quark mesons is small and may be 
ignored.  One must instead use the four-quark condensates directly; this is
the proceedure we will follow.  As the temperature dependence of these is not 
generally known, we will of course be using the temperature dependence 
as it comes from our theory.

The remaining non-perturbative quantity that needs to be specified is the gluon 
condensate.  At finite temperature, we may write for the gluon condensate
\beq
\langle G^2\rangle\simeq\langle:G^2:\rangle + \langle G^2\rangle^{P.T.}_T
\label{gluegen}
\eeq
where the perturbative term $\langle G^2\rangle^{P.T.}_T$ has been calculated
by Kapusta~\cite{kap} and found to be 
\beq
\langle G^2\rangle^{P.T.}_T\simeq\frac{8\pi^2}{15}T^5 ,
\label{gluept}
\eeq
which at $T_c\simeq 200$ MeV is a very small fraction of the zero-temperature
nonperturbative term, 
$\langle G^2\rangle^{P.T.}_{T_c}<\frac{1}{10}\langle:G^2:\rangle$.
This leads to our assumption, also made in Ref.~\cite{dn,fhl}, that the 
temperature dependence
of $\langle :G^2:\rangle_T$ is weak. In the present work we assume that 
\beq
\langle G^2\rangle_T\simeq\langle G^2\rangle .
\label{gluet}
\eeq
This also means that the gluon condensate is nonvanishing for temperatures 
$T>T_c$, so that the vacuum is still nonperturbative.

\subsection{QCD Sum Rules for Meson Masses}
\hspace{.5cm}

In this subsection, we present the results for the QCD sum rules for meson
masses~\cite{rry1} extended to finite temperature using the Matsubara
formalism~\cite{m}.  In the next subsection we will introduce similar sum 
rules, with however some important differences, for the four-quark condensates.

The light-quark mesons that have been succesfully treated in QCD sum rules
include the lightest that couple to a given $J^{\Gamma}$ as specified
in Eq.~(\ref{current}).  Generally, these
mesons come in nearly degenerate isospin doublets.  For the light-quark
mesons corresponding to the elementary Dirac operators, we have
the following mesons to consider:  the f$_0$(1370)-a$_0$(1450)~\cite{kgv}
in the 0$^{++}$ scalar channel ($\Gamma$=1); the 
$\rho$(776)-$\omega$(783)~\cite{svz} in the 1$^{--}$ vector channel 
($\Gamma =\gamma_\mu$); a$_1$(1260)-f$_1$(1285)~\cite{rry2} in 
the 1$^{++}$ axial vector channel ($\Gamma =\gamma_5\gamma_\mu$).  The 
$\pi$(140)-$\eta$(549) in the 0$^{-+}$ psudoscalar channel ($\Gamma=\gamma_5$), 
is an exception for which QCD sum rules are no good.  In the two scalar
channels, the f$_0$ and a$_0$ have a different character in that only the 
f$_0$ couples directly to the 0$^{++}$ scalar glueball, but this
does not affect our calculation of the ${\bar q}q$ part of the correlator.

The correlators $\Pi^\Gamma (p)$ generally have a rich tensoral structure,
and it is sufficient for our purposes to consider only one of these, which
we will refer to as $\Pi^{\prime\Gamma}$.  In the case of the vector, axial 
vector, and scalar mesons $\Pi^{\prime\Gamma}$ is related to 
$\Pi^{\Gamma}$ of Eq.~(\ref{eq-pi}) respectively by 
$\Pi^{\prime v_\|}=\pi p^{-2}g^{\mu 0} g^{\nu 0}\Pi^{v}_{\mu\nu}$ for 
the longitudinal correlator of the 
$\rho$(776)-$\omega$(783); $\Pi^{\prime av}=-\frac{\pi}{3} g^{\mu\nu} 
\Pi^{av}_{\mu\nu}$ for the a$_1$(1260)-f$_1$(1285); and 
$\Pi^{\prime s}=\pi \Pi^{s}$ for the f$_0$(1370)-a$_0$(1450).
Noting the near-degeneracy of the I=0,1 mesons of each type except the $ps$,
one can simplify the theory considerably by assuming an exact 
isospin degeneracy; in this case the correlators are 
multiplied by a factor of two relative to the value in the absence of isospin 
symmetry~\cite{rry1}.  

We next characterize the correlators as they arise from the quark current, 
coming back to the glueball contribution of the 0$^{++}$ mesons later.  A 
convenient summary of the various ingredients may be found in the 
review~\cite{rry1}, except for the extension to finite temperature, which 
follows the arguments found above.  

The L.H.S. of the correlators arising from the quark current have the 
following form in momentum for the $s, v$, and $av$ cases,
\beq
\Pi^{\prime\Gamma}_{L.H.S}(Q^2,T)=a_\Gamma I_n(-p^2)+ b_\Gamma
 \frac{
\langle\frac{\alpha}{\pi}G^2\rangle}{(-p^2)^{n(\Gamma,b)}}
+c_\Gamma\frac{{\hat{\hat Q}}^\Gamma}{(-p^2)^{n(\Gamma,c)}} .
\label{lhsmmsr}
\eeq
In these expressions, $-p^2=Q^2$ is the Euclidian momentum, and the function 
$I_n(-p^2)$ characterizes the perturbative quark terms (see Fig.~2(a))
and has the following form
\beq
I_n(-p^2)&\equiv & i_n(-p^2) 
\label{capI1}
\eeq
for the $s$ and $av$ cases, and
\beq
I_n(-p^2)\equiv i_0(-p^2)&+&\frac{2}{-p^2}\int^\infty_0d\omega^2n_F
(\frac{\omega}{2T})\nonumber\\
& &-\frac{2}{3(-p^2)}\int^\infty_{4m_\pi^2}d\omega^2n_B
(\frac{\omega}{2T}).
\label{capI2}
\eeq
for the $v_\|$ case. Here $n_B(z)=(e^z-1)^{-1}$, $n_F(z)=(e^a+1)^{-1}$, and
\beq
i_n(-p^2)\equiv \frac{1}{n!}\int^{S_0}_0d\omega^2\frac{\omega^{2n}}{-p^2+
\omega^2}\tanh(\omega/4T) .
\label{lcI}
\eeq
The last integral in Eq.~(\ref{capI2}) describes the damping due to thermal 
excitation of pions~\cite{bs} and is particular to the longitudinal
channel.
We have assumed that the contribution of the continuum to the dispersion 
relation begins at $s=S_0$, and that this piece of the R.H.S. is transferred 
to the L.H.S. of the sum rule, where it regulate the large-$s$ contributions 
of the perturbative terms.  For this reason, the integrals in Eq.~(\ref{lcI}) 
run only up to $S_0$.  This corresponds to the usual treatment of the 
continuum at $T=0$.  

The following light-quark mesons, which have slightly more complicated 
currents, have also been successfully treated by QCD sum rules:  The 
D$_1$(1235)~\cite{rry2} in the axial 1$^{+-}$ channel 
($J^{(1^{+-})}_{\mu}=i{\bar q} \gamma_5 \stackrel{\leftrightarrow}{
\partial}_\mu q$);
the a$_2$(1320)-f$_2$(1270)~\cite{as} in the tensor 2$^{++}$ channel 
($J^{(2^{++})}_{\mu\nu}=i{\bar q} (\gamma_\mu \stackrel{\leftrightarrow}
{\partial}_\nu+
\gamma_\nu \stackrel{\leftrightarrow}{ \partial}_\mu)q$);
and the H$_2$(1670)~\cite{as} in the axial tensor 2$^{-+}$ channel
($J^{(2^{-+})}_{\mu\nu}=i{\bar q} (\gamma_\mu \gamma_5\stackrel{\leftrightarrow}
{\partial}_\nu+\gamma_\nu \gamma_5 \stackrel{\leftrightarrow}{\partial}_\mu)q$.

The correlator $\Pi^{\prime\Gamma}$ for the axial meson has the same
form as given in Eq.~(\ref{lhsmmsr}).  For the tensor and axial tensor mesons, 
$\Pi^{\prime\Gamma}$ has the form
\beq
\Pi^{\prime\Gamma}_{L.H.S}(Q^2,T)=a_\Gamma I_n(-p^2)+ b_\Gamma (-\log (-p^2))
\langle\frac{\alpha}{\pi}G^2\rangle 
+c_\Gamma\frac{{\hat{\hat Q}}^\Gamma}{(-p^2)^{n(\Gamma,c)}} .
\label{lhstat}
\eeq
The following three ${\hat {\hat Q}}$ correspond to the axial (1$^{+-}$),
tensor (2$^{++}$), and axial tensor (2$^{-+}$) mesons respectively:
${\hat{\hat Q}}^{a}={\hat Q}^{ps}-\frac{1}{9}|J_0|^2$,
${\hat{\hat Q}}^{t}={\hat Q}^{v}$, and ${\hat{\hat Q}}^{at}={\hat Q}^{av}$,
where we have introduced rather obvious abbreviations for these channels.
The appropriate $I_n(-p^2)$ is that of Eq.~(\ref{capI1}) in these cases.

The coefficients of Eqs.~(\ref{lhsmmsr},\ref{lhstat}) are given in Table~1.
We refer the reader to the original literature~\cite{svz, rry1, rry2, kgv}
for the details not given here.  

To obtain the sum rule, we make the Borel transform
\beq
\Pi^{\prime\Gamma}(M_B^2,T)\equiv 
{\hat L}_{M_B}\Pi^{\prime\Gamma}(Q^2,T) = \lim_{Q^2,n\rightarrow\infty, 
Q^2/n=M_B^2}\frac{(Q^2)^{n+1}}{n!}(\frac{d}{d(Q^2)})^2\Pi^
{\prime\Gamma}(Q^2,T))
\label{bt}
\eeq
of both the R.H.S. and left-hand side.   After making the Borel transform of 
the meson-mass sum rules, we may write the L.H.S. of the sum rule in the 
following generic fashion
\beq
\Pi^{\prime\Gamma}_{L.H.S}(M_B^2,T)=A_\Gamma M_B^{2(n+1)} e_n(M_B) + 
B_\Gamma\frac{\langle\frac{\alpha}{\pi}G^2\rangle} {M_B^{2n(\Gamma,B)}}
+C_\Gamma\frac{{\hat{\hat Q}}^\Gamma}{M_B^{2n(\Gamma,C)}} ,
\label{lhsmsr}
\eeq
 The function 
$e_n(M_B)$ is given in all cases by $M_B^{2(n+1)}e_n(M_B)\equiv 
{\hat L}_{M_B}I_n(Q^2)$, with
\beq
{\hat L}_{M_B}i_n(Q^2) \equiv \frac{1}{n!}\int^{S_0}_0d\omega^2\omega^{2n}
e^{-\frac{\omega^2}{M^2_B}}\tanh(\frac{\omega}{4T}) .
\label{e}
\eeq
Note, in the case of the longitudinal channel~\cite{bs} for the vector meson, we
have from Eqs.~(\ref{capI1}) and (\ref{capI2})
\beq
M_B^2e_0(M_B) \equiv \int^{S_0}_0d\omega^2
e^{-\frac{\omega^2}{M^2_B}}\tanh(\frac{\omega}{4T})+2\int^\infty_0 d\omega^2
n_F(\frac{\omega}{2T}) -
\frac{2}{3}\int^\infty_{4m_\pi^2}d\omega^2n_B(\frac{\omega}{2T}) .
\label{etwit}
\eeq
The coefficients $A$, $B$, $C$, and the various $n$'s, are given in 
Table~2 for all the cases.  

In the case of the $0^{++}$ meson, we follow the philosophy of Ref.~\cite{kgv}
in which the scalar meson is treated as an hybird. i.e., the current for the
$0^{++}$ case is
\beq
     J^{0^{++}} = \beta J^s +(1-|\beta|)J^{GB(0^{++})},
\label{mixed}
\eeq
where $J^s$ is the scalar quark current defined in Eq.~(\ref{current}) and
$J^{GB(0^{++})}$ is the scalar glueball current. Details are given in 
Ref.~\cite{kgv}.  Our approach
differs from that of Ref.~\cite{rry1,rry2} because the mesons considered there
correspond to the f$_0$(980) and a$_0$(980) mesons, which are not ${\bar q} q$
mesons.  Instead, we consider the $0^{++}$ to be the f$_0$(1370) and 
a$_0$(1450) mesons.  Because these mesons are hybrid, there is in addition 
to the quark current and the correlator arising from the coupling of the 
${\bar q} q$ pair to the glueball, giving
rise to a ``glueball" contribution
\beq
\Pi^{\prime Glueball}_{L.H.S}=4\pi M_B^6(\frac{\alpha_S}{\pi})^2
+8\pi^3(\frac{\alpha_S}{\pi})^2
\Gamma^{(6)}+(\frac{\alpha_S}{\pi})^3\frac{8\pi^5\Gamma^{(8)}}{M_B^2} .
\label{glueball}
\eeq
To obtain the full correlator for the scalar meson, we combine this with 
the quark contribution in Table~2,
\beq
\Pi^{\prime 0^{++}}_{L.H.S}=\beta^2\Pi^{\prime quark,0^{++}}_{L.H.S}+
(1-|\beta|)^2\Pi^{\prime glueball}_{L.H.S.} ,
\label{smeson}
\eeq
where the value $\beta=0.7$ reproduces the mass of the $M^{0^{++}}$=1370 MeV.
Note that the cross term, proportional to $\beta(1-|\beta|)$, does not appear 
in the expression in Eq.~(\ref{smeson}) because this term vanishes under the 
Borel transform in the form we use.

Our correlator does not agree exactly with the one found in Ref.~\cite{kgv}.
As in the case of the other mesons, we multiply the quark term by 
a factor of two to take into account that we are mixing taking the isoscalar 
and isovector mesons to be degenerate.  Additionally, we use an 
{\it unsubtracted} dispersion relation and regulate the strong dependence 
on $M_B$ in the glueball similar to the regulation of the continuum for 
quarks.  Since the glueball makes a relatively small contribution for this 
meson, these differences are relatively unimportant numerically.

To complete the sum rule for the meson masses, we need to specify the 
expression for the phenomenological correlator, $\Pi^{\prime\Gamma}_{R.H.S.}$.
In all cases, $\Pi^{\prime\Gamma}_{R.H.S.}(M_B^2,T)$ is proportional to 
$e^{-m_\Gamma^2/M_B^2}$, where $m_\Gamma$ is
the mass of the meson created by current $J^\Gamma$.  The meson-mass sum
rule is obtained by equating the the logarithmic derivitive of the R.H.S.
and L.H.S.,
\beq
\frac{\partial_{M_B^{-2}}\Pi^{\prime\Gamma}_{R.H.S.}}
{\Pi^{\prime\Gamma}_{R.H.S.}}(M_B^2,T)
 =\frac{\partial_{M_B^{-2}}\Pi^{\prime\Gamma}_{L.H.S.}}
{\Pi^{\prime\Gamma}_{L.H.S.}}(M_B^2,T).
\label{srs}
\eeq
According to the theory, the sum rule is valid only in a region there the 
combination in Eq.~(\ref{srs}) is relatively insensitive to $M_B$, which in 
practice means within a region where the sum rule has a plateau.  Note that
the logarithmic derivitive of the R.H.S. is just the square of the meson 
mass, and for this reason it is not necessary to know the constant of 
proportionality in $\Pi^{\prime\Gamma}_{R.H.S.}(M_B^2,T)$. Later, when we 
discuss the four-quark sum rules the constant of proportionality on the R.H.S. 
is quite important, and we will carefully discuss it for these cases.

We give some of the results for $m_\Gamma$ in Table~3.  We also give the
values of the threshold parameters $S_0^\Gamma$ and the factorized values of
the four-quark condensates appropriate for $T=0$.  As $T$ is increased, we
cannot solve the QCD sum rules for meson masses without knowing the 
temperature dependence of the
four-quark condensates.  In Sect.~3 we present two models for this.

\subsection{Four-Quark Condensate Sum Rules}
\hspace{.5cm}

We next want to look for a sum rule whose solution gives the four-quark 
condensates themselves.  The correlators for the four-quark sum rules will
be the correlators $\Pi^\Gamma$(p,T) in Eq.~(\ref{eq-pit}) for the 
currents $J^\Gamma$ corresponding to the five elementary Dirac structures.
The correlators are in most instances not the same as they were for 
the meson-mass sum rules.  For example, the sum rules for $\Pi^{\sigm}$ and
$\Pi^{ps}$ are going to be considered in this context.  We will give our
arguments in the next section why the use of the psudoscalar correlator
the four-quark condensate may be justified while it is not justified for 
$m_\pi$, as noted earlier.  

One can easily verify that $J^{\sigm}$ couples to the 1$^{--}$ channel from the
fact that a nonvanishing antisymmetric tensor for the overlap in 
Eq.~(\ref{eq-eta}) can be constructed from the vector meson 
$|m(p,\lambda)\rangle$ intermediate state in $\Pi^{\prime\sigm}_{R.H.S.}$ as 
\beq
\langle 0|\sigma_{\mu\nu}|m(p,\lambda)\rangle =g_{\sigm} (p_\mu 
\varepsilon^{(\lambda)}_\nu (p)-p_\nu \varepsilon^{(\lambda)}_\mu (p)) ,
\label{sig}
\eeq
where $\varepsilon^{(\lambda)}_\mu$ is the polarization vector for the vector
meson.  Because the current $J^{\sigm}$ couples to the same 1$^{--}$ channel as 
the $\rho$-$\omega$, the $\sigm$ channel is generally not considered as an 
independent case.  

To identify the quantity that we will be solving the sum rules to obtain,
consider first the R.H.S of $\Pi^\Gamma (p)$ for the case of $T=0$.  In 
this case, we write for the residue at the pole in Eq.~(\ref{imrho})
\beq
\sum_{\lambda}\langle 0|J^\Gamma|m_\alpha ({\vec p},\lambda)\rangle
\langle m_\alpha ({\vec p},\lambda)|J^\Gamma|0\rangle = C F^\Gamma (\alpha )
P^\Gamma_\alpha({\vec p})
\label{residue0}
\eeq
where $P^\Gamma_\alpha({\vec p})=\sum_\lambda \Psi({\vec p},\lambda)
\Psi^\dagger ({\vec p},\lambda)$ is a projection operator whose form is 
specified by the quantum numbers for the particular meson in question, $C$ is 
a overall scale common to all mesons $\Gamma$, and where $F^\Gamma (\alpha )$ 
is a function that characterizes the differences among the different mesons and
whose form is not needed in this work.  Explicit forms suggested for these 
functions may be found in Ref.~\cite{bs,kkp,gkl,w}.
We give $P^\Gamma (p)$
along with the definitions of the correlators $\Pi^{\prime\Gamma}(Q^2,T)$ for 
each of the five elementary Dirac operators in Table~4.  Note that for the 
vector meson, we use the transverse correlator, 
$\Pi^{\prime v\bot}=-\pi g^{\mu\nu}\Pi^{v}_{\mu\nu}$.  

We will be making some model assumptions about the spectrum of states 
$|m_\alpha\rangle$ that enter Eq.~(\ref{residue0}), one being to replace the 
continuum by a set of resonances having four-momentum 
($|E_\alpha |\equiv \sqrt{{\vec p}^2+m_\alpha^2},{\vec p}$).  In this
case, consider the matrix element $\langle 0|J^\Gamma J^\Gamma |0\rangle$
and insert a complete set of states $|m_\alpha\rangle$ between the two currents
to obtain
\beq
\langle 0|J^\Gamma J^\Gamma |0\rangle &=& \sum_{\lambda\alpha}\int
\frac{d^3{\vec p}_\alpha}{(2\pi )^3} \frac{1}{2 E_\alpha }
\langle 0|
J^\Gamma |m_\alpha ({\vec p},\lambda)\rangle\langle m_\alpha ({\vec p},\lambda)|
J^\Gamma|0\rangle \nonumber\\
&=& C \kappa^\Gamma
\label{residue1}
\eeq
where
\beq
\kappa^\Gamma \equiv \sum_{\alpha} F^\Gamma(\alpha )\int
\frac{d^3{\vec p}_\alpha}{(2\pi )^3} \frac{{\check P}_\alpha^\Gamma(p)}
{2 E_\alpha} ,
\label{kappa}
\eeq
where ${\check P}^\Gamma_\alpha$ denotes contraction over the Dirac indices of
$P^\Gamma_\alpha$.  The value of $\kappa^\Gamma$ is dependent on various
cutoff factors that are not known, so we will eventually fix the value
of $\kappa^\Gamma$ phenomenologically to assure that the four-quark
condensates attain their factorized value for $T=0$.  We may 
solve Eq.~(\ref{residue1}) for the scale $C$ to 
obtain 
\beq
C=\langle 0|J^\Gamma J^\Gamma |0\rangle /\kappa^\Gamma .
\label{C}
\eeq

Now, using Eqs.~(\ref{imrho}) and  (\ref{C}) we may write the sum in 
Eq.~(\ref{residue0}) as
\beq
\rho(E,{\vec p})=\frac{\langle 0|J^\Gamma J^\Gamma |0\rangle}{\kappa^\Gamma}
\int\frac{d^3{\vec p}_\alpha}{(2\pi )^3}\sum_{\alpha} F^\Gamma (\alpha )
P^\Gamma_\alpha({\vec p})
\frac{(2\pi)^3\delta({\vec p}-{\vec p}_\alpha)}{2E_\alpha}
\delta(E-E_\alpha) .
\label{rhof}
\eeq
Note that with the help of Eq.~(\ref{rhof}) we are able to express the R.H.S.
of the correlator $\Pi^\Gamma (p)$ in terms of the quantity 
$\langle 0|J^\Gamma J^\Gamma |0\rangle$ that is closely related to the 
four-quark condensate $\4qc$.  Such an expression will enable us to find a
set of sum rules for the four-quark condensates, independent of the
QCD sum rules for meson masses.

The extension of Eq.~(\ref{rhof}) to the case of finite temperature is 
straightforward using the results of Sect.~(2.2).  We noted there that 
the R.H.S. for the case of finite temperature was similar to the R.H.S.
at $T=0$ except for the appearance of a temperature-dependent occupation
probability.  Since the quantity appearing on the R.H.S. is a meson, the
correct occupation probability is $\coth(\frac{E_\alpha}{2T})$.  Then,
\beq
\rho_T(E,{\vec p})=\frac{\langle\langle 0|J^\Gamma J^\Gamma |0\rangle\rangle}
{\kappa^\Gamma(T)}
\int\frac{d^3{\vec p}_\alpha}{(2\pi )^3}\sum_{\alpha} F^\Gamma (\alpha )
P^\Gamma_\alpha({\vec p})\coth(\frac{E_\alpha}{2T})
\frac{(2\pi)^3\delta({\vec p}-{\vec p}_\alpha)}{2E_\alpha}
\delta(E-E_\alpha) ,
\label{rhoft}
\eeq
where
\beq
\kappa^\Gamma(T) \equiv \sum_{\alpha} F^\Gamma(\alpha )\int
\frac{d^3{\vec p}_\alpha}{(2\pi )^3} \frac{1}{2 E_\alpha}
P_\alpha^\Gamma(p) \coth^2(\frac{E_\alpha}{2T}).
\label{kappat}
\eeq

In the QCD sum rule for meson masses, the higher-lying states that contribute
to the dispersion relation (i.e., the continuum, assumed to begin at $s=S_0$) 
are 
transferred to the L.H.S. of the sum rule, where they regulate the perturbative 
term at large momentum.  We make the assumption
that the continuum plays exactly the same role for the four-quark condensate 
sum rules, and therefore we characterize the continuum by exactly the
same functions $I_n(M_B^2)$ that characterize the meson-mass sum rules.  Here, 
we parametrize $S_0$ to be a function of temperature, the functional form 
carrying the burden of describing the dropping of the threshold $S_0$ as the
temperature is raised, as well as the temperature dependence of $\kappa (T)$
in Eq.~(\ref{kappat}).  With these asumptions, the R.H.S. of $\Pi^\Gamma(p)$
has the following form,
\beq
\Pi^\Gamma_T(\omega,{\vec p};T)=\frac{P^\Gamma (p)}{-p^2+m_\Gamma^2}
\frac{\langle\langle 0|J^\Gamma J^\Gamma |0\rangle\rangle}
{\kappa^\Gamma}\coth(m_\Gamma/2T) .
\label{rhs4qsr}
\eeq

The left-hand side of $\Pi^\Gamma (p,T)$ is expanded in an o.p.e. as before in 
the case of meson
masses.  The gluon condensate is 
treated as in the case of the meson-mass QCD sum rule.  The correlator
$\Pi^{\prime\Gamma}_{L.H.S.}(Q^2,T)$ has the same form as given in 
Eq.~(\ref{lhsmmsr}), with the treatment of the
continuum as given in the preceeding paragraph.  There is a glueball
effect in the $s$ channel now only in $\rho_T(E,{\vec p})$, since the current 
is purely of the form given in Eq.~(\ref{current}).  The values of the 
coefficients are given in Table~5.

The temperature dependence of the
four-quark condensate $\4qc$ is treated as follows.  We would like to relate
it to the quantity $\langle\langle 0|J^\Gamma J^\Gamma |0\rangle\rangle$ 
that appears on
the R.H.S. of the sum rule in order to obtain a set of equations to determine
it.  This connection is found by making a Fierz rearrangement of the quark
operators appearing in $\4qc$, so we write
\beq
\4qc =\sum_{\Gamma^{\prime}} c_{\Gamma^{\prime}}(\Gamma )
\langle\langle 0|J^{\Gamma^\prime} J^{\Gamma^\prime} |0\rangle\rangle
\label{fierz}
\eeq
making use of the following relationship among the Gell-Mann matrices
\beq
t^a_{\alpha\gamma}t^a_{\lambda\beta}=2\delta_{\alpha\beta}\delta_{\gamma\lambda}
-\frac{2}{3}\delta_{\alpha\gamma}\delta_{\lambda\beta} .
\label{gellmann}
\eeq
After making the Fierz rearrangement, the four-quark condensates 
${\hat{\hat Q}}^\Gamma$ are related to ${\hat Q}^\Gamma$ as shown in Table~6.
We have taken $|J_0|^2={\hat Q}^v$, which although not an exact result
differs from it only by terms that vanish in the factorization approximation.  
This is the only place where we make a factorization assumption
(other than at $T=0$).

To obtain the four-quark sum rule, we make the Borel transform as
given in Eq.~(\ref{bt}) of both the R.H.S. and L.H.S. and then equate them,
\beq
\Pi^{\prime\Gamma}_{L.H.S.}(M_B^2,T)=\Pi^{\prime\Gamma}_{R.H.S.}(M_B^2,T) .
\label{4qsr}
\eeq
As in the case of the meson-mass sum rule, the equality is possible only in
a region where the sum rule is rather insensitive to $M_B$.
The R.H.S. and L.H.S. of the four-quark sum rules are given in Tables~7
and 8, respectively.  The L.H.S. correlators have the same form as given
in Eq.~(\ref{lhsmsr}).  Note that these constitute five equations in the five 
unknowns $\4qc$ and other quantities.  We will come back to the solution
of this equation later in the next section.

\section{Toward a Theory of Four-Quark Condensates at Finite Temperature}
\hspace{.5cm}

In this section, we will present two models of the four-quark condensates and
the corresponding meson masses.  For our first model, we combine the results 
of Sect.~(2.3) with a parametrization of the four-quark condensates and the 
continuum.  

\subsection{Model I:  Parametrizing the Condensates and Continuum}
\hspace{.5cm}

We first look at a simplified model in which we use the meson mass QCD sum 
rules in the
$v_\|$, $av$, $s$, $a$, $at$, and $t$ channels, assuming that the temperature 
dependence of the four-quark condensates behaves according to
\beq
{\hat{Q}}^\Gamma (T)={\hat{Q}}^\Gamma (0)F(T) .
\label{qhh}
\eeq
For this simplified model, we will be assuming that the {\it same} function 
$F(T)$ describes the temperature dependence of all four-quark condensates.  
This amounts to a factorization assumption.  For the present calculation, 
we take
\beq
F(T)= 1-T^2/T_c^2 ,
\label{F}
\eeq
which corresponds to the mean-field assumption of Ref.~\cite{fhl}, where the
temperature dependence of the quark condensate was taken to be 
proportional to the squar-root of this quantity.  We take 
$T_c=220$ MeV~\cite{bs}.  In our second model, discussed in Sect.~3.2, we will 
be able to check factorization for $T\neq T_c$.   

Our second assumption regards the behavior of the temperature dependence of the
continuum, which we are taking to begin at $s=S_0^\Gamma (T)$.  As we are unable
to calculate the temperature dependence of the continuum from first 
principles, we parametrize this quantity.  We have found that we get
reasonable and consistent solutions of our sum rules using all our models
if we take
\beq
S_0^\Gamma (T)=S_0^\Gamma (0)G(T) .
\label{thresh}
\eeq
with
\beq
G(T)=1-T^2/\Lambda^2,~~{\rm with}~\Lambda=331~MeV ,
\label{G}
\eeq
with $S_0(0)$ given in Table~3.
As a result of this parametrization, our results depend on the value of the
cutoff parameter $\Lambda$.  With this choice of $S_0^\Gamma (T)$, 
the continuum moves down toward the resonance $\Gamma$ as the temperature
increases.  Such behavior is expected to occur in a realistic system, to
the extent that all the masses drop and hadronic resonances get increasing
widths with increasing temperature.

Using these two assumptions, and the QCD sum rule specified by Eq.~(\ref{srs}),
we get equations for $m_\Gamma^2(T)$ of the following form,
\beq
m_\Gamma^2(T)=H^\Gamma (T,S_0^\Gamma(T),M_B^\Gamma,{\hat {\hat Q}}^\Gamma (T)) .
\label{m1}
\eeq
The L.H.S. of the correlators we use are given in Table~2.  Note that
in this model, ${\hat {\hat Q}}^\Gamma$ is proportional to $F(T)$ with the
values at $T=0$ given in Table~3.  The solution of 
these equations is taken in the region where $m_\Gamma$ is insensitive to 
$M_B^\Gamma$.  We have found that to a reasonable approximation 
for the $av$, $s$, and $a$ cases this region occurs where $M_B^\Gamma$ attains 
its $T=0$ value.   However, for the $v_\|$, $at$, and $t$ cases, the region of
stability in $M_B^\Gamma$ drops with increasing $T$.  A simple, approximate
description obtains by taking $M_B^{v_\|}\approx m_{v_\|}(T)+.5$ GeV$^2$,
$M_B^{at}\approx m_{at}(T)$, and $M_B^{t}\approx m_{t}(T)$.  

The results are shown in Fig.~3.  
\begin{figure}
\begin{center}
\epsfig{file=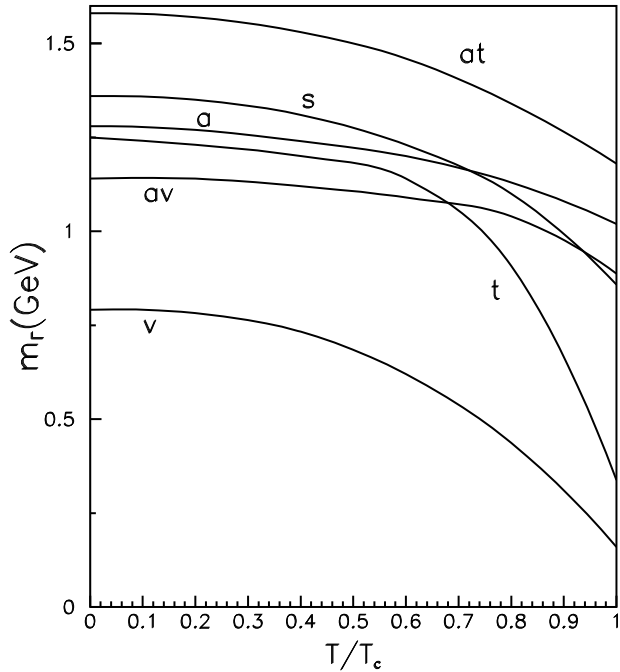,width=10cm}
\caption{\label{Fig. 3} Solutions of Eq.~(\ref{m1}) assuming factorized
forms for four-quark condensates.}
\end{center}  
\end{figure}             
As in Ref.~\cite{bs}, the mass of the vector meson drops 
to zero at $T\approx T_c$, but the rate of decrease is not nearly as fast here.
The behavior of the tensor meson is particularly interesting.  Between 
$T\approx 3T_c/5$ and $T\approx T_c$, $m^2_t(T,M_B^t )$ changes 
its character several times:  from a function whose relevant region of 
stability is an asymptote at large $M_B^t$, to a function whose region of 
stability is a broad shoulder, and finally to a function having as its only 
stable solution one at relatively small values of $M_B^t$.  We are able to 
express the complicated change in the location of stable regions by 
$M_B^t\approx m_t(T)$.  The axial tensor meson is also an interesting case.
The relevant region of $M_B^{at}$ is a broad shoulder at all $T$.  The region 
of stability tends to move to smaller $M_B^{at}$, corresponding to smaller 
$m_{as}$,  as $T$ increases toward $T=T_c$.  Additionally, in this limit
the correlator becomes increasingly sensitive to the condensates.

It is easily verified from Fig.~3 and Table~3 that the gap between the 
continuum and the meson pole decreases as expected when $T$ approaches 
$T_c$ for all mesons except the tensor meson.  In this case, the drop of the 
meson mass is more rapid than the dropping of $S_0(T)$.  Note that the Borel 
transform becomes less effective at suppressing features of the excited states 
that are not modelled in the theory as the separation between the resonance 
and the continuum decreases.  Thus, in this model as well as the next, the 
theoretical uncertainties grow as $T_c$ is approached in a fashion that
is increasingly dependent on the unknown parameter $\Lambda$ in Eq.~(\ref{G}).
These uncertainties may be able to be decreased somewhat by introducing
specific forms of $F^\Gamma (\alpha )$ and $P^\Gamma_\alpha ({\vec p})$
given in Eq.~(\ref{residue0}) or by extrapolating to results obtained at
asymptotically large $T$, where uncertainties may be again better
controlled~\cite{kj}.

Although there is considerable theoretical uncertainty in our results in
the vacinity of $T=T_c$, we note the following interesting feature that may 
hint at new physics.  Starting at $T\approx T_c/2$ the tensor meson mass
begins to drop quickly, and the rapid drop continues as $T$ approaches
$T_c$.  Except for this case, and that of the vector meson, the meson masses 
remain relatively large at the transition temperature.  The rapid decrease of 
the tensor meson mass appears to be associated with the unusual behavior of 
the corresponding correlator at large $T$, noted above. Because of this, the 
tensor meson may be particularly sensitive to the temperature dependence of 
the four-quark condensate and be an important case deserving careful 
experimental study.

\subsection{Model II:  Solving the Four-Quark Condensate Sum Rules}
\hspace{.5cm}

In this section, we remove one of the assumptions of the previous subsection,
namely, we no longer parametrize the four-quark condensates but rather 
determine these condensates self-consistently using the four-quark QCD sum
rules whose R.H.S. and L.H.S. are given respectively in Tables~7 and 8.
Note that these four-quark sum rules depend on the masses of the mesons in the 
$v$, $av$, $s$, $ps$, and $\sigm$ channels.  In deriving these sum rules,
we have already taken the meson mass on the R.H.S. to be the one corresponding
to the lightest meson for each current.  Thus, we take the mass $m_v(T)$ to be 
that 
corresponding to the $\rho - \omega$ meson; $m_{av}(T)$ to be that 
corresponding to the $a_1 - f_1$ meson; and $m_s(T)$ to be that 
corresponding to the $a_0-f_0$ meson.  Since $J^{\sigm}$ couples to the 
1$^{--}$ channel, we take the mass in the $\sigm$ channel to be $m_v(T)$, 
corresponding to the $\rho - \omega$ meson.  For the mass of the pion, we 
take the free-space value $m_\pi =140$ MeV following the observation in the 
model of Ref.~\cite{hk} that the pion mass is rather stable with temperature 
up to $T_c$.  The values of $m_s(T)$, $m_v(T)$, and $m_{av}(T)$ are of course 
taken to be solutions of the corresponding three meson-mass sum rules 
discussed already in Model I, above.

We may write the theory as follows.  Note that the four-quark condensate sum 
rules may be solved algebraically for the five matrix elements
$\langle\langle 0| J^\Gamma J^\Gamma |0\rangle\rangle$ if Eq.~(\ref{fierz})
is used to express ${\hat {\hat Q}}^\Gamma$ in terms of 
$\langle\langle 0| J^\Gamma J^\Gamma |0\rangle\rangle$.  The result may
be written 
\beq
\langle\langle 0| J^\Gamma J^\Gamma |0\rangle\rangle (T)=
K^\Gamma (T,m(T),S_0(T),M_B^{\prime\Gamma}) ,
\label{4q}
\eeq
i.e., as an explicit function of a Borel mass $M_B^{\prime\Gamma}$ for each
matrix element $\langle\langle 0| J^\Gamma J^\Gamma |0\rangle\rangle$,
the three relevant masses ($m(T)$), and the five threshold functions 
($S_0(T)$).  The remaining three equations are of course the three meson-mass 
sum rules of Eq.~(\ref{m1}),
\beq
m_\Gamma^2(T)=H^\Gamma (T,m_\Gamma,S_0^\Gamma(T),M_B^\Gamma,
{\hat {\hat Q}}^\Gamma (T))~~{\rm for}~\Gamma=s,~v,~{\rm and}~av .
\label{m2}
\eeq
To get a completely self-consistent set of equations, we again express
${\hat {\hat Q}}^\Gamma$ in terms of 
$\langle\langle 0| J^\Gamma J^\Gamma |0\rangle\rangle$ using Eq.~(\ref{fierz}).

Our Model II thus consists of eight coupled nonlinear equations for the
eight unknowns, consisting of the five matrix elements 
$\langle\langle 0| J^\Gamma J^\Gamma |0\rangle\rangle$ and the three meson 
masses $m_s(T)$, $m_v(T)$, and $m_{av}(T)$.  As we are unable to
calculate the temperature dependence of the continuum from first principles,
we continue to parametrize this as in Eq.~(\ref{thresh}) and Eq.~(\ref{G}),
above.  We take the thresholds $S_0^\Gamma$ to be the same as those in the 
$T=0$ meson-mass sum rules, Table~3, except in the case of the $ps$ and 
$\sigm$ channels, where we take $S_0^{\sigm} = S_0^v$ and $S_0^{ps}=1300$ MeV.
The parameters $\kappa^\Gamma$ in 
Eq.~(\ref{kappat}) are obtained by solving the four-quark condensate sum 
rules at $T=0$ and adjusting the values of these
parameters to obtain the known {\it factorized} values of the matrix elements 
$\langle 0|J^\Gamma J^\Gamma |0\rangle$.  Note that we are assuring that the
quark condensates attain their free-space values by our choice of 
$\kappa^\Gamma$.  The values that assure this are given in Table~9.

\begin{figure}
\begin{center}
\epsfig{file=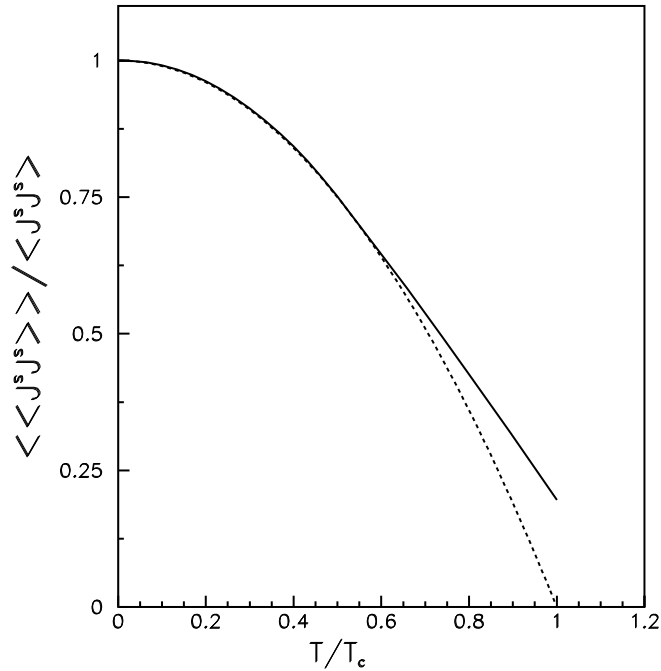,width=10cm}
\caption{Temperature-dependence of $R(T)$ defined in Eq.~(\ref{RR}) 
(solid line) for the
scalar four-quark condensate compared to temperature dependence of four-quark
condensates in the factorized, mean-field approximation (dashed line).}
{\label{Fig.4}}                                                               
\end{center}  
\end{figure}             

To obtain numerical solutions of the nonlinear system given in Eqs.~({\ref{4q},
\ref{m2}), we begin at $T=0$, where we know that the 
solution of our equations gives the well-known results for the meson masses 
and (factorized) four-quark condensates.  We then slowly increase the 
temperature, letting $T\rightarrow \delta$, solving by iteration the
eight equations in eight unknowns until a stable solution is reached for the
for the new meson masses and condensates at $T=\delta$.  Then, we increase 
the temperature $T\rightarrow T+\delta$ and iterate again for a stable 
solution.  The process is continued up to the transition temperature $T_c$.
Our results do not rule out the existence of multiple solutions at finite
temperature.  We have in all cases presented the solutions that evolve
continuously from the solutions at $T=0$.

\begin{figure}
\begin{center}
\epsfig{file=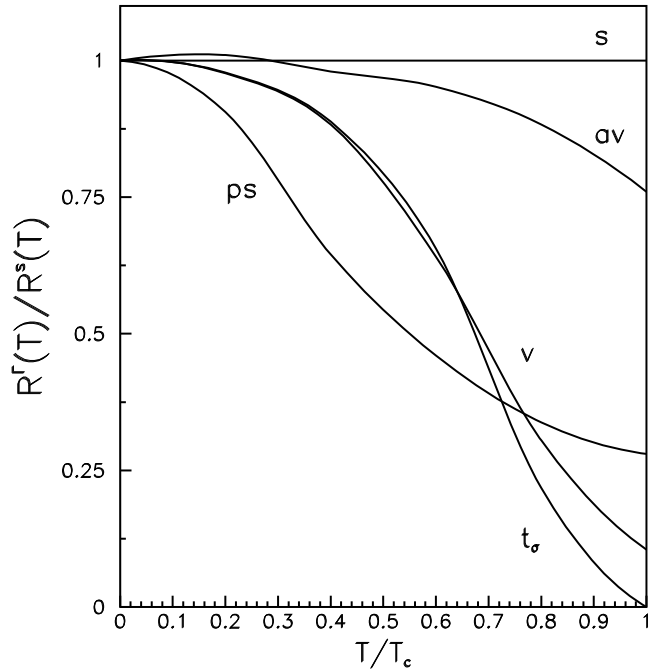,width=10cm}
\caption{T-dependence of QCD sum rule determined four-quark 
condensates as a function of temperature.  Note that the T-dependence
is given relative to the the scalar four-quark condensate shown in Fig.~4.}
{\label{Fig.5}}                                                               
\end{center}  
\end{figure}             

We obtain the solution of each sum rule by finding the region of stability 
as a function of the Borel mass for each quantity of interest.  We thus
find a Borel mass for each of the eight unknown quatities at each $T$.  Once 
we know
the condensates, we can solve the sum rules in the $a$, $t$, and $at$ 
channels.  Clearly, the meson masses in these three channels are determined as 
a consequence of the eight coupled equations, but the corresponding mesons do 
not appear as part of the four-quark condensates sum rule.  

We give the numerical solutions for the coupled equations next.  We will
denote the the temperature dependence of the condensates by $R^\Gamma (T)$, 
so that
\beq
\langle\langle 0|J^\Gamma J^\Gamma |0\rangle\rangle =R^\Gamma (T) 
\langle 0| J^\Gamma J^\Gamma |0\rangle .
\label{RR}
\eeq
\begin{figure}
\begin{center}
\epsfig{file=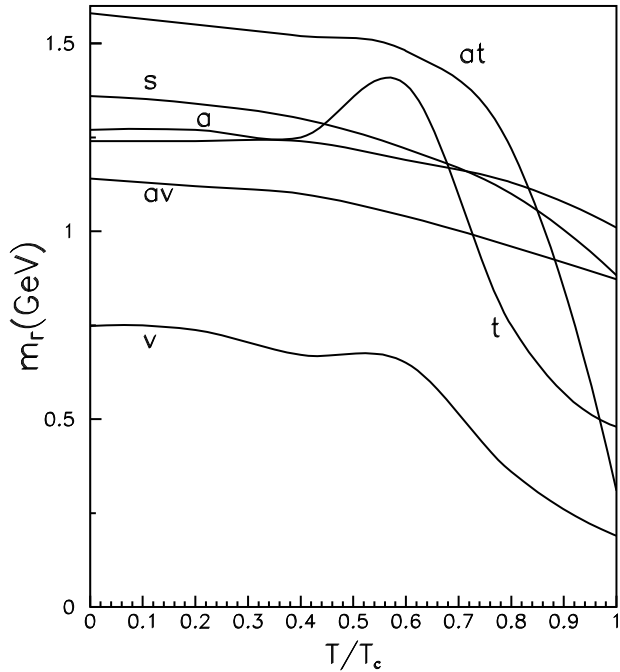,width=10cm}
\caption{Meson masses as a function of temperature using our QCD sum rule
determination of the four-quark condensates.}
{\label{Fig.6}}                                                               
\end{center}  
\end{figure}             
First, in Fig.~4, we show the behavior $R^\Gamma (T)$ for the four-quark scalar 
condensate.  
The solid curve shows the solution to our coupled equations, and
the dashed line shows for comparison the mean-field temperature dependence
given in Eq.~(\ref{F}).  We see that the scalar four-quark condensate falls 
off with temperature very much as expected from factorized 
approximations~\cite{fhl}.  We have found that the details of the temperature
dependence in Fig.~4 is governed by our choice of $G(T)$ in Eq.~(\ref{G}).

Next, consider the relative temperature dependence of all of the four-quark 
condensates.  These are shown in Fig.~5.
We see here that the scalar condensate has the slowest fall-off with $T$.
Additionally, it is clear that all condensates $\Gamma$ fall off in 
$T$ with different rates; the ratio 
$R^\Gamma (T)/R^s(T)$ is much less sensitive to $G(T)$ than 
$R^\Gamma(T)$ itself.  Vacuum saturation would imply the same rate of 
fall-off, so 
we see that even at a modest temperature of $T\approx T_c/2$ there is a 
considerable violation of factorization.  
                                                  
At $T$=0 the factorization approximation of the four-quark condensates is
also in question. In a study of the ratio of the isovector hadronic  to the
muon pair production in $e^+e^-$ data~\cite{gbp} using a sum rule that
weights the heavier states more than the usual sum rule~\cite{svz}, a value of
the vector four-quark condensate is
larger than the factorized value. Using this value we find that we get
a satisfactory fit to the rho-meson mass. Moreover,
in the present work, we observe that the
$T=0$, $B=0$ sum rules give all meson masses consistent with experiment for
our parameterization with $T=0$ factorization. We have explored the 
additional
features of four-quark condensates for finite $T$, and there is no 
difficulty
for us to modify our results to include correction factors for $T=0$.
                                                                    
We give the numerical solutions for the meson masses in Fig.~6.  
Note that the temperature dependence of the meson masses for all cases except
the tensor and axial tensor mesons are quite similar to those of Model I
shown in Fig.~3.  Because the tensor and axial tensor show more sensitivity to 
the four-quark condensates, these cases may provide particularly interesting
candidates to explore experimentally.

We have noted that accuracy of the correlator in the psudoscalar channel 
for the mass of the $\pi$ meson is very poor.  Novikov, et al.~\cite{nsvz}
have indicated that for channels that couple easily to the vacuum, such 
as the psudoscalar channel, instanton corrections identified as ``direct
instantons" may be particularly important.  The subject of the instanton
contributions to meson mass correlators has studied in some detail
in a recent review~\cite{shu}.  Here different models of correlators have 
been compared, showing how instantons can improve the
correlator out to large separations $\tau$.
Although direct instantons presumably also contribute to the psudoscalar and
other four-quark condensate sum rules that we have used, we have 
normalized these sum rules to the factorized value of the four-quark
condensate at $T=0$.  The fact that we are only relying on the four-quark sum 
rules to provide the temperature dependence of the condensates and not their
$T=0$ values presumably improves the reliability of the sum rule 
for our purpose.

\section{Summary and Conclusions}
\hspace{.5cm}

This paper has focused on the physics of the four-quark condensates
${\hat Q}^\Gamma$, which constitute fundamental, nonperturbative features of 
quantum chromodynamics needed for understanding how hadron masses
(especially those of the light-quark mesons), as well as interactions among 
hadrons, arise.  These as well as the other condensates all govern essential 
details 
of the phase transition that is believed to have occured as the early universe 
expanded and cooled from the era dominated by the quark-gluon plasma into
the next, in which the quarks and gluons had coalesced into the familiar 
hadrons.  Understanding the physics of this phase transition has been 
identified as one of the opportunities that presents itself to nuclear and 
particle physics community at the next generation of experimental facilities 
of RHIC and perhaps the LHC.  

In spite of the importance of ${\hat Q}^\Gamma\equiv \4qc$, these quantities are 
currently quite poorly understood at both the phenomenological and at the 
theoretical levels.  For example in the literature, the temperature 
dependence of ${\hat Q}^\Gamma$ is almost universally described through a 
factorized, mean-field approximation in which 
${\hat Q}^\Gamma \propto \qq^2\propto 1-T^2/T_c^2$, with $T_c\approx 200$ MeV.  
In such a formulation, the dominant nonperturbative quark contribution to the
masses of light-quark mesons is the quark condensate $\qq$.  In this paper, we 
have proposed a dynamical theory for the temperature dependence of 
${\hat Q}^\Gamma$ that generalizes the familiar gap equation for the quark 
condensate $\qq$.  In our theory, the masses of the light-quark mesons are 
practically independent of $\qq$; instead these masses depend on four-quark 
condensates that do not factorize for $T\neq 0$.

Our work is based on finite-temperature QCD sum rules in the Matsubara 
formalism and requires a model of the hadronic spectrum.  The model of the 
hadronic spectrum we have chosen is the discretized model adopted in 
Ref.~\cite{bs}.  It contains, in addition, a parameter that characterizes the 
temperature-dependence of the continuum.  Equations are obtained coupling 
light-quark meson masses $m_\Gamma$ and $\4qc$.  Stable solutions are found 
in which $\4qc$ all decrease as $T$ increases toward $T_c$, but at different 
rates, so that 
factorization becomes increasingly violated the farther $T$ is varied away 
from $T=0$.  

We stress the connection between the temperature dependence of meson masses 
and the fundamental four-quark condensates in hot quark matter.  The
tensor 2$^{++}$ mesons a$_2$(1320)-f$_2$(1270) are particularly important
for giving quantative information on the four-quark condensates.
The temperature dependence of meson masses constitute predictions 
of our theory that may at least in principle be checked experimentally by 
measurement of final electromagnetic states ($\gamma\gamma$, $e^+e^-$, or 
$\mu^+\mu^-$) or perhaps semi-electromagnetic states such as $\pi\gamma$.

Our results do not depend
on achieving thermodynamic equilibrium in relativistic heavy ion collisions,
but only local thermodynamic equilibrium.  We also note that the 
temperature- (and more generally, density-) dependence of meson masses 
may affect predictions of transport models modelling the evolution of matter
into and out of the quark-gluon plasma.  Results such as the ones we give may
be incorporated in such models able to follow the evolution of the 
local temperature. 

Acknowledgement:  The authors thank H. Forkel and T. Steele for enlightening
discussions.  MBJ acknowledges the hospitality and support of L. Kisslinger and 
the Physics Department of Carnegie Mellon University 
during the academic year 1997-98, when this work was begun.  This work was 
also performed under the auspices of the U.S. Department of Energy under 
contract W-7405-ENG-36 and the National Science Foundation under grant 
PHY-9722143.

\newpage


\begin{table}
\begin{center}
\caption{Meson mass correlator $\Pi^{\prime\Gamma}_{L.H.S.}(Q^{2},T)$.
Coefficients correspond to Eq.~(19) for the $av$, $s$, and $v_\|$ cases,
and to Eq.~(23) for the $a$, $t$, and $at$ cases.  For the $s$ case,
there is also a glueball contribution, given in the text.} 
\vskip 0.25in
\renewcommand{\arraystretch}{1.5}
\begin{tabular}{ccccccc}
\hline\hline
Case $\Gamma$ & $n$ & $a_{\Gamma}$ & $n(\Gamma,b)$ & $b_{\Gamma}$ &
$n(\Gamma,c)$ & $c_{\Gamma}$ \cr
\hline
$av$ & 1 & ${1\over 4\pi}\left(1+{\alpha_{s}\over\pi}\right)$ & 1 &
$\llap{--}{\pi\over 12}$ & 2 & $2\pi^{2}\alpha_{s}$ \cr
$s$ & 1 & ${3\over 8\pi}\left(1+{11\over 3}{\alpha_{s}\over\pi}\right)$ & 1 &
${\pi\over 8}$ & 2 & $\pi^{2}\alpha_{s}$ \cr
$v_\|$ & 0 & ${1\over 4\pi}\left(1+{\alpha_{s}\over\pi}\right)$ & 2 &
${\pi\over 12}$ & 3 & $\llap{--}2\pi^{2}\alpha_{s}$ \cr
$a$ & 1 & ${1\over 8\pi}\left(1+{\alpha_{s}\over\pi}\right)$ & 1 &
$\llap{--}{\pi\over 24}$ & 2 & $4\pi^{2}\alpha_{s}$ \cr
$t$ & 2 & ${3\over 5\pi}\left(1-{\alpha_{s}\over\pi}\right)$ & --- &
$\llap{--}{8\pi\over 9}$ & 1 & $\llap{--}4\pi^{2}\alpha_{s}$ \cr
$at$ & 2 & ${3\over 5\pi}\left(1-{\alpha_{s}\over\pi}\right)$ & --- &
$\llap{--}{8\pi\over 9}$ & 1 & $\llap{--}4\pi^{2}\alpha_{s}$ \cr
\hline\hline
\end{tabular}
\end{center}
\end{table}


\begin{table}
\begin{center}
\caption{Borel transform of meson mass correlator 
$\Pi^{\prime\Gamma}_{L.H.S}(M_{B}^{2},T)$.  Coefficients correspond to Eq.~(25)
in all cases.} 
\vskip 0.25in
\renewcommand{\arraystretch}{1.5}
\begin{tabular}{ccccccc}
\hline\hline
Case $\Gamma$ & $n$ & $A_{\Gamma}$ & $n(\Gamma,B)$ & $B_{\Gamma}$ &
$n(\Gamma,C)$ & $C_{\Gamma}$ \cr
\hline
$av$ & 1 & ${1\over 4\pi}\left(1+{\alpha_{s}\over\pi}\right)$ & 0 &
$\llap{--}{\pi\over 12}$ & 1 & $2\pi^{2}\alpha_{s}$ \cr
$s$ & 1 & ${3\over 8\pi}\left(1+{11\over 3}{\alpha_{s}\over\pi}\right)$ & 0 &
${\pi\over 8}$ & 1 & $\pi^{2}\alpha_{s}$ \cr
$v_\|$ & 0 & ${1\over 4\pi}\left(1+{\alpha_{s}\over\pi}\right)$ & 1 &
${\pi\over 12}$ & 2 & $\llap{--}\pi^{2}\alpha_{s}$ \cr
$a$ & 1 & ${1\over 8\pi}\left(1+{\alpha_{s}\over\pi}\right)$ & 0 &
$\llap{--}{\pi\over 24}$ & 1 & $4\pi^{2}\alpha_{s}$ \cr
$t$ & 2 & ${3\over 5\pi}\left(1-{\alpha_{s}\over\pi}\right)$ & \llap{--}1 &
$\llap{--}{8\pi\over 9}$ & 0 & $\llap{--}4\pi^{2}\alpha_{s}$ \cr
$at$ & 2 & ${3\over 5\pi}\left(1-{\alpha_{s}\over\pi}\right)$ & \llap{--}1 &
$\llap{--}{8\pi\over 9}$ & 0 & $\llap{--}4\pi^{2}\alpha_{s}$ \cr
\hline\hline
\end{tabular}
\end{center}
\end{table}


\begin{table}
\begin{center}
\caption{Results for meson-mass sum rules, $T=0$.  Factorized values of the
four-quark condensates are also given.}
\vskip 0.25in
\renewcommand{\arraystretch}{1.5}
\begin{tabular}{cccc}
\hline\hline
Case $\Gamma$ & $S_0^\Gamma$ (MeV$^{2}$) & $m_{i}$ (MeV) & 
$<{\hat {\hat Q}}^\Gamma>_{fact} /<{\bar q}q>^2$ \cr
\hline
$av$ & 1700   & 1140  &$-\frac{176}{81}$ \cr
$s$ & 3000 & 1360 &$-\frac{176}{27}$ \cr
$v$ & 1500  & 748 &$~\frac{112}{81}$ \cr
$a$ & 2740  & 1240 &$-\frac{20}{81}$ \cr
$t$ & 2500  & 1270 &$-\frac{16}{9}$ \cr
$at$  & 3500  & 1580 &$~\frac{16}{9}$ \cr
\hline\hline
\end{tabular}
\end{center}
\end{table}


\begin{table}
\begin{center}
\caption{Definition of projection operator $P^\Gamma (p)$ and four-quark 
condensate correlator $\Pi^{\prime\Gamma}(Q^{2},T)$.}
\vskip 0.25in
\renewcommand{\arraystretch}{1.5}
\begin{tabular}{ccc}
\hline\hline
Case $\Gamma$ & $P^{\Gamma}(p)$ & Def.\ $\Pi^{\prime\Gamma}$ \cr
\hline
$av$ & ${1\over m_{av}^{2}}\left(p_{\mu}p_{\nu}-p^{2}g_{\mu\nu}\right)$ &
$-{\pi p^{2}g^{\mu\nu}\over 6}\Pi_{\mu\nu}^{av}$  \cr
$s$ & 1 & $\pi\Pi^{s}$ \cr
$\sigm$ & ${1\over m_{\sigm}^{2}}(p_{\mu}p_{\lambda}g_{\nu\rho}+
p_{\nu}p_{\rho}g_{\mu\lambda}$ & 
${\pi P^{\mu\nu,\rho\lambda}\over 12}\Pi_{\mu\nu,\rho\lambda}^{\sigm}$ \cr

 & $~~~~-p_{\mu}p_{\rho}g_{\nu\lambda}-p_{\nu}p_{\lambda}g_{\mu\rho})$ & 
  \cr

$ps$ & \llap{--}1 & $\pi\Pi^{ps}$  \cr
$v_{\bot}$ & ${1\over m_{v}^{2}}\left(p_{\mu}p_{\nu}-p^{2}g_{\mu\nu}\right)$ & $-
{\pi g^{\mu\nu}\over 3}\Pi_{\mu\nu}^{v}$ \cr
\hline\hline
\end{tabular}
\end{center}
\end{table}


\begin{table}
\begin{center}
\caption{Four-quark condensate correlator $\Pi^{\prime\Gamma}_{L.H.S.}(Q^2,T)$.
Coefficients corresond to Eq.~(19).} 
\vskip 0.25in
\renewcommand{\arraystretch}{1.5}
\begin{tabular}{ccccccc}
\hline\hline
Case~$\Gamma$ & $n$ & $a_{\Gamma}$ & $n(\Gamma,b)$ & $b_{\Gamma}$ &
$n(\Gamma,c)$ & $c_{\Gamma}$ \cr
\hline
$av$ & 2 & ${1\over 4\pi}\left(1+{\alpha_{s}\over\pi}\right)$ & --- &
0 & 1 & \llap{--}$\pi^{2}\alpha_{s}$ \cr
$s$ & 1 & ${3\over 8\pi}\left(1+{11\over 3}{\alpha_{s}\over\pi}\right)$ & 1 &
${\pi\over 8}$ & 2 & $\pi^{2}\alpha_{s}$ \cr
$\sigm$ & 1 & $\llap{--}{1\over 8\pi}$ & 1 &
${\pi\over 3}$ & 2 & $\llap{16}\pi^{2}\alpha_{s}$ \cr
$ps$ & 1 & ${3\over 8\pi}\left(1+{11\over 3}{\alpha_{s}\over\pi}\right)$ & 1 &
${\pi\over 8}$ & 2 & $\pi^{2}\alpha_{s}$ \cr
$v_\bot$ & 1 & ${1\over 4\pi}\left(1+{\alpha_{s}\over\pi}\right)$ & 1 &
\llap{--}${\pi\over 12}$ & 2 & \llap{2}$\pi^{2}\alpha_{s}$ \cr
\hline\hline
\end{tabular}
\end{center}
\end{table}


\begin{table}
\begin{center}
\caption{Fierz rearranged four-quark condensates.} 
\vskip 0.25in
\renewcommand{\arraystretch}{1.5}
\begin{tabular}{ccccccc}
\hline\hline
Case $\Gamma$  & ${\hat {\hat Q}}^{\Gamma}$ & $C_{v}$ & $C_{av}$ & $C_{s}$ 
& $C_{ps}$ &
$C_{t}$ \cr
\hline
$v$ & $\hat{Q}^{av}+{2\over 9}\vert J_{0}\vert^{2}$ & ${29\over 27}$ 
& \llap{--}${5\over 9}$ & ${14\over 9}$ & ${14\over 9}$ & 0 \cr
$av$ & $\hat{Q}^{v}+{2\over 9}\vert J_{0}\vert^{2}$ & ${11\over 27}$ &
${11\over 9}$ & \llap{--}${22\over 9}$ & \llap{--}${22\over 9}$ & 0 \cr
$s$ & $\hat{Q}^{t}+{2\over 3}\vert J_{0}\vert^{2}$ & ${2\over 9}$ &
${2\over 3}$ & \llap{--}${22\over 3}$ & ${14\over 3}$ & ${1\over 3}$ \cr
$\sigm$ & $\hat{Q}^{s}+{1\over 36}\vert J_{0}\vert^{2}$ 
& $\llap{--}{53\over 108}$ & ${19\over 36}$ & \llap{--}${11\over 9}$ 
& ${4\over 9}$ & $\llap{--}{1\over 4}$ \cr
$a$ & $\hat{Q}^{ps}-{1\over 9}\vert J_{0}\vert^{2}$ & ${25\over 54}$ &
$\llap{--}{11\over 18}$ & \llap{--}${5\over 18}$ & ${25\over 18}$ 
& $\llap{--}{1\over 4}$ \cr
$ps$ & $\hat{Q}^{at}+{2\over 3}\vert J_{0}\vert^{2}$ & ${2\over 9}$ &
${2\over 3}$ & \llap{--}${22\over 3}$ & ${14\over 3}$ & ${1\over 3}$ \cr
$t$ & $\hat{Q}^{v}$ & ${1\over 3}$ & 1 & \llap{--}2 & \llap{--}2 & 0 \cr
$at$ & $\hat{Q}^{av}$ & 1 & ${1\over 3}$ & 2 & 2 & 0 \cr
\hline\hline
\end{tabular}
\end{center}
\end{table}


\begin{table}
\begin{center}
\caption{Four-quark correlator $\Pi^{\prime\Gamma}_{R.H.S.}(M_B^2,T)$.}
\vskip 0.25in
\renewcommand{\arraystretch}{1.5}
\begin{tabular}{cc}
\hline\hline
Case $\Gamma$ & $\Pi^{\prime\Gamma}_{R.H.S.}(M_{B}^{2},T)$ \cr
\hline
$av$ & ${\langle\langle 0\vert J^{av}J^{av}\vert 0\rangle\rangle\over\kappa^{av}}
{m_{av}^{2}\over 2}\pi\,\coth \left( {m_{av}\over 2\tau} \right) 
e^{-m_{av}^{2}/M_{B}^{2}}$ \cr
$s$ & ${\langle\langle 0\vert J^{s}J^{s}\vert 0\rangle\rangle\over\kappa^{s}}
\pi\,\coth \left({m_{s}\over 2\tau}\right) e^{-m_{s}^{2}/M_{B}^{2}}$ \cr
$\sigm$ & \llap{--}${\langle\langle 0\vert J^{\sigm}J^{\sigm}\vert 
0\rangle\rangle\over\kappa^{\sigm}}
\pi\,\coth \left({m_{v}\over 2\tau}\right) e^{-m_{v}^{2}/M_{B}^{2}}$ \cr
$ps$ & \llap{--}${\langle\langle 0\vert J^{ps}J^{ps}\vert 
0\rangle\rangle\over\kappa^{ps}}
\pi\,\coth \left({m_{ps}\over 2\tau}\right) e^{-m_{ps}^{2}/M_{B}^{2}}$ \cr
$v_\bot$ & \llap{--}${\langle\langle 0\vert J^{v}J^{v}\vert 
0\rangle\rangle\over\kappa^{v}}
\pi\,\coth \left({m_{v}\over 2\tau}\right) e^{-m_{v}^{2}/M_{B}^{2}}$ \cr
\hline\hline
\end{tabular}
\end{center}
\end{table}


\begin{table}
\begin{center}
\caption{Four-quark correlator $\Pi^{\prime\Gamma}_{L.H.S.}(M_B^2,T)$.} 
\vskip 0.25in
\renewcommand{\arraystretch}{1.5}
\begin{tabular}{ccccccc}
\hline\hline
Case $\Gamma$ & $n$ & $A_{\Gamma}$ & $n(\Gamma,B)$ & $B_{\Gamma}$ &
$n(\Gamma,C)$ & $C_{\Gamma}$ \cr
\hline
$av$ & 2 & ${1\over 4\pi}\left(1+{\alpha_{s}\over\pi}\right)$ & --- &
0 & 0 & \llap{--}$\pi^{2}\alpha_{s}$ \cr
$s$ & 1 & ${3\over 8\pi}\left(1+{11\over 3}{\alpha_{s}\over\pi}\right)$ & 0 &
${\pi\over 8}$ & 1 & $\pi^{2}\alpha_{s}$ \cr
$\sigm$ & 1 & $\llap{--}{1\over 8\pi}$ & 0 &
${\pi\over 3}$ & 1 & $\llap{16}\pi^{2}\alpha_{s}$ \cr
$ps$ & 1 & ${3\over 8\pi}\left(1+{11\over 3}{\alpha_{s}\over\pi}\right)$ & 0 &
${\pi\over 8}$ & 1 & $\pi^{2}\alpha_{s}$ \cr
$v_\bot$ & 1 & ${1\over 4\pi}\left(1+{\alpha_{s}\over\pi}\right)$ & 0 &
\llap{--}${\pi\over 12}$ & 1 & \llap{2}$\pi^{2}\alpha_{s}$ \cr
\hline\hline
\end{tabular}
\end{center}
\end{table}


\begin{table}
\begin{center}
\caption{Values of $\kappa^{\Gamma}$ and factorzied values of $<0|{\bar q}
\Gamma q{\bar q} \Gamma q|0>$.} 
\vskip 0.25in
\renewcommand{\arraystretch}{1.5}
\begin{tabular}{ccc}
\hline\hline
Case $\Gamma$ & $\kappa^{\Gamma}$ (GeV$^{2}$) & $\langle 0\vert : \overline{q}\Gamma
q\,\overline{q}\Gamma q : \vert 0\rangle/\langle 0 \vert :\overline{q}q:\vert 0\rangle^{2}$ 
 \cr
\hline
$av$ & 0.0021$\phantom{0}$ & ${1\over 3}$  \cr
$s$ & 0.00073 & ${11\over 12}$  \cr
$\sigm$ & 0.028$\phantom{00}$ & \llap{--}1 \cr
$ps$ & \llap{--}0.00034 & ${1\over 12}$ \cr
$v_\bot$ & 0.0036$\phantom{0}$ & \llap{--}${1\over 3}$  \cr
\hline\hline
\end{tabular}
\end{center}
\end{table}


\end{document}